# On the Essence of Lagrange's Equations

Peng SHI[1, 2 *]

[1] Logging Technology Research Institute, China National Logging Corporation, No.50 Zhangba five road, Xi'an, 710077, P. R. China

[2] Well Logging Technology Pilot Test Center, China National Logging Corporation, No.50 Zhangba five road, Xi'an, 710077, P. R. China

*Corresponding author. Email: sp198911@outlook.com;

***Abstract.*** From a new perspective, this paper rederives Lagrange’s equations. By applying the chain rule of differentiation, the intrinsic relationship between the momentum theorem and the kinetic energy theorem is first established. Subsequently, expressing the differential form of energy conservation in an arbitrary coordinate system and performing suitable differential operations yields Lagrange’s equations. Generalized forces and generalized displacements are shown to be component representations of forces and displacements in a chosen coordinate system. Consequently, the essence of Lagrange’s equations is identified as the transformation of the kinetic energy theorem into the momentum theorem via the chain rule for composite functions, thereby revealing how energy conservation constructs momentum conservation.

## 1. Introduction

Lagrangian mechanics offers a reformulation of classical mechanics that provides profound insight into dynamical systems. It significantly simplifies the analysis of systems subject to multiple constraints [1, 2], and is widely extended and applied in other disciplinary fields [3-8], such as multi-rigid-body systems and robotics, vibration analysis and modal analysis, as well as continuum mechanics and field theory. Deeply analyzing the physical essence of Lagrangian mechanics holds significant theoretical and practical value.

Based on the form of Lagrange's equations, Lagrangian mechanics articulates, through the "language of the theorem of kinetic energy," dynamical laws that are equivalent to Newton's second law. In classical mechanics, the theorem of momentum and the theorem of kinetic energy are generally regarded as two independent dynamical laws. However, Lagrangian mechanics naturally suggests that these two theorems may actually describe the same dynamical characteristics, merely expressed from different perspectives. To date, Lagrange's equations have been derived independently from two physical principles: d'Alembert's principle and Hamilton's principle [1]. Consequently, both d'Alembert's principle and Hamilton's principle are regarded as the physical foundations of Lagrangian mechanics. Nevertheless, regarding the fundamental dynamical question of why the theorem of kinetic energy can be used to derive the theorem of momentum, no clear answer has yet been provided. Given the diversity of energy forms in matter and the fact that material interactions can be expressed as energy transfer and conversion, deriving Lagrange's equations from the perspective of the non-independence of the work-energy theorem and the momentum theorem, and exploring its essence, is of great significance to analytical mechanics and all of physics.

In this study, I rederives Lagrange's equations from a new perspective. First, the intrinsic relationship between the theorem of momentum and the theorem of kinetic energy is revealed through the chain rule of differentiation. Then, by expressing the differential form of the energy conservation law in an arbitrary coordinate system and performing appropriate differential operations, Lagrange's equations are derived. The results show that generalized forces and generalized displacements in Lagrangian mechanics are the component representations of forces and displacements in a specific coordinate system. The essence of Lagrange's equations lies in converting the kinetic energy theorem into the momentum theorem via the chain rule for differentiating composite functions, leveraging the intrinsic relationship between the two theorems.

## 2. Particle kinematics in different perspectives

Considering the motion of a particle under the action of a given force, there exists a one-to-one correspondence between its spatial position and time throughout its trajectory. Consequently, the velocity and acceleration of the particle can be equivalently described either as functions of its spatial position or as functions of time. To draw an analogy between these alternative descriptions of particle motion and the frameworks used in continuum mechanics [9], this paper defines the representation of kinematic quantities as functions of time as the Lagrangian description, and their representation as functions of spatial coordinates as the Eulerian description. While this definition may initially seem unconventional, its validity will become evident through the subsequent analysis of the relationship between theorem of momentum and the theorem of kinetic energy.

Although the physical quantities themselves remain unchanged when described within these different frameworks, their mathematical representations differ significantly. To clearly distinguish between these two forms of expression, the following convention is adopted unless otherwise stated: lowercase letters denote physical quantities in the Lagrangian description, whereas uppercase letters denote those in the Eulerian description. For specific commonly used physical quantities, further distinctions will be clarified through explanatory definitions in the text.

In Lagrangian perspective, the velocities of a particle at two infinitesimally close instants of time satisfy the following relationship:

$$\boldsymbol{u}(t+\delta t)=\boldsymbol{u}(t)+\frac{d\boldsymbol{u}}{dt}\delta t \tag{1}$$

here, $\boldsymbol{u}$ is the velocity vector in Lagrangian perspective, $t$ is the time, $\delta t$ is the time increment, and $d/dt$ is the time derivative. In Eulerian perspective, the velocities of a particle at two points in space corresponding to two infinitely close instants of time satisfy the following relationship:

$$\boldsymbol{U}(\boldsymbol{x}+\delta\boldsymbol{x})=\boldsymbol{U}(\boldsymbol{x})+\delta\boldsymbol{x}\cdot\nabla\boldsymbol{U} \tag{2}$$

here, $\boldsymbol{x}$ is the displacement vector, $\delta\boldsymbol{x}$ is the displacement of a particle during $\delta t$, $\boldsymbol{U}$ is the velocity vector in Eulerian perspective which is a function with respect to displacement vector, $\nabla$ is the gradient operator.

Since the following relationship holds for a certain particle:

$$\boldsymbol{u}(t) = \boldsymbol{U}(\boldsymbol{x}(t)) = \frac{\delta \boldsymbol{x}}{\delta t} \tag{3}$$

The acceleration expression of the particle described in Lagrangian perspective and in Eulerian perspective are obtained and expressed as:

$$\frac{d\boldsymbol{u}}{dt} = \boldsymbol{U} \cdot \nabla \boldsymbol{U} = \frac{1}{2}\nabla\left(U^2\right) + \nabla \times \boldsymbol{U} \times \boldsymbol{U} \tag{4}$$

here, $U$ is the module of $\boldsymbol{U}$.

## 3. Particle dynamics in Lagrangian and Eulerian perspectives

In classical mechanics, the following law holds:

$$\boldsymbol{f}\,\delta t = m\delta \boldsymbol{u} \tag{5}$$

here, $\boldsymbol{f}$ is the resultant force on an object, $m$ is the mass of the object which is a constant. Equation (5) is termed the theorem of momentum. Multiplying both sides of equation (5) by velocity $\boldsymbol{u}$ yields the following equation:

$$\boldsymbol{F} \cdot \delta \boldsymbol{x} = m\boldsymbol{U} \cdot \nabla \boldsymbol{U} \cdot \delta \boldsymbol{x} \tag{6}$$

here, $\boldsymbol{F}$ the resultant force in Eulerian perspective.

Since $\boldsymbol{U}$ is parallel to $\delta\boldsymbol{x}$, which can be obtained from Equation (3), The following equation is obtained:

$$\boldsymbol{U} \times \left(\nabla \times \boldsymbol{U}\right) \cdot \delta \boldsymbol{x} = 0 \tag{7}$$

Then, Equation (8) is rewritten as follows:

$$\boldsymbol{F} \cdot \delta \boldsymbol{x} = \frac{1}{2} m\nabla\left(U^2\right) \cdot \delta \boldsymbol{x} \tag{8}$$

Equation (8) represents the kinetic energy theorem. Since Equation (8) is obtained by performing appropriate vector operations on Equation (5), this shows that the theorem of momentum and the theorem of kinetic energy are not independent fundamental principles, but rather two equivalent integral forms of Newton's second law, describing the cumulative effect of force from the perspectives of time and space, respectively.

## 4. Derivation of Lagrange's Equations

Above, physical quantities such as displacement vectors, velocities, and external forces were described using vectors. Here, we express these physical quantities in a certain coordinate system.

In a given coordinate system, displacement can be expressed as [10]:

$$\boldsymbol{x} = q^i \boldsymbol{g}_i = q_i \boldsymbol{g}^i \tag{9}$$

With

$$\boldsymbol{g}_i = \frac{\partial \boldsymbol{x}}{\partial q^i}, \boldsymbol{g}^i = \frac{\partial \boldsymbol{x}}{\partial q_i} \tag{10}$$

here, $q^i$ is the contravariant components, $q_i$ is the covariant component, $\boldsymbol{g}_i$ is the covariant basis and $\boldsymbol{g}^i$ is the contravariant basis. Then, the differential of displacement the and the differential of velocity can be respectively expressed as:

$$d\boldsymbol{x} = \frac{\partial \boldsymbol{x}}{\partial q^i} dq^i = \frac{\partial \boldsymbol{x}}{\partial q_i} dq_i \tag{11}$$

$$d\dot{\boldsymbol{x}} = d\boldsymbol{u} = d\dot{q}^i \boldsymbol{g}_i \tag{12}$$

With Equations (11) and (12) and considering velocity as a composite function, the following relation is obtained:

$$\boldsymbol{g}_i = \frac{\partial \boldsymbol{x}}{\partial q^i} = \frac{\partial \boldsymbol{u}}{\partial \dot{q}^i} \tag{13}$$

In the coordinate system, the theorem of kinetic energy of a particle is expressed as:

$$Q_i dq^i = m\frac{d}{dt}\left(\frac{d}{dt}\left(q_i \boldsymbol{g}^i\right)\right)\cdot dq^i \boldsymbol{g}_i \tag{14}$$

with $Q_i$ the covariant component of resultant force in the coordinate system. Rearranging Equation (14), it can be rewritten as:

$$\left(m\frac{d}{dt}\left(\frac{d}{dt}q_i \boldsymbol{g}^i\right)\cdot \boldsymbol{g}_i - Q_i\right)dq^i = 0 \tag{15}$$

Considering now the relation:

$$m\frac{d}{dt}\left(\frac{d}{dt}q_i \boldsymbol{g}^i\right)\cdot \boldsymbol{g}_i = m\frac{d}{dt}\left(\frac{d}{dt}\left(q_i \boldsymbol{g}^i\right)\cdot \boldsymbol{g}_i\right) - m\frac{d}{dt}\left(q_i \boldsymbol{g}^i\right)\cdot \frac{d\boldsymbol{g}_i}{dt} \tag{16}$$

Submitting Equation (13) into Equation (16), Equation (16) is rewritten as:

$$m\frac{d}{dt}\left(\frac{d}{dt}q_i \boldsymbol{g}^i\right)\cdot \boldsymbol{g}_i = m\frac{d}{dt}\left(\boldsymbol{u}\cdot\frac{\partial \boldsymbol{u}}{\partial \dot{q}^i}\right) - m\boldsymbol{u}\cdot\frac{\partial \boldsymbol{u}}{\partial q^i} \tag{17}$$

Therefore, Equation (15) is rewritten as:

$$\left(\frac{d}{dt}\left(\frac{\partial T}{\partial \dot{q}^i}\right) - \frac{\partial T}{\partial q^i} - Q_i\right)dq^i = 0 \tag{18}$$

here, $T = \frac{1}{2}mU^2$ is the kinetic energy.

When the resultant force acting on the particle is a conservative force, the resultant force can be expressed in terms of a scalar potential function as:

$$\boldsymbol{F} = \nabla V = \frac{\partial V}{\partial q_i}\boldsymbol{g}^i = Q_i \boldsymbol{g}^i \tag{19}$$

Submitting Equation (19) into Equation (18) and having $L=T+V$, then Equation (18) is expressed as:

$$\left(\frac{d}{dt}\left(\frac{\partial L}{\partial \dot{q}^i}\right) - \frac{\partial L}{\partial q^i}\right)dq^i = 0 \tag{20}$$

with $L=T+V$. Equation (20) is identical to Lagrange's equations. Here the definition of a conservative force is slightly different from the traditional formulation. Under the new definition, $L$ is the mechanical energy which is conserved quantity. Therefore, the essence of Lagrangian mechanics lies in exploiting the non-independence between energy conservation and momentum conservation to construct momentum conservation by means of energy conservation.

**Conclusion**

In this paper, Lagrange's equations have been rederived from a novel perspective that emphasizes the relationship between momentum and kinetic energy theorems. By applying the chain rule of differentiation, we have shown that the theorem of momentum and the theorem of kinetic energy are not independent but intrinsically related. This relationship forms the theoretical basis for the derivation of Lagrange's equations. Expressing the differential form of the energy conservation law in an arbitrary coordinate system, followed by appropriate vector operations, directly yields Lagrange's equations. This demonstrates that Lagrange's equations can be obtained without invoking d'Alembert's principle or Hamilton's principle as a starting point. Generalized forces and generalized displacements in Lagrangian mechanics are shown to be merely the component representations of actual forces and displacements in a chosen coordinate system. This clarifies the coordinate-dependent nature of these quantities. The fundamental nature of Lagrange's equations is identified as the transformation of the kinetic energy theorem into the momentum theorem via the chain rule for composite functions. In other words, Lagrange's equations construct momentum conservation through energy conservation.

## Acknowledgments

This work was supported by Deep Earth Probe and Mineral Resources Exploration–National Science and Technology Major Project (2024ZD1003000) and the scientific research and technology development projects of China National Petroleum Corporation (2026ZS002).

## Statements and Declarations

The author declares that he has no known competing financial interests or personal relationships that could have appeared to influence the work reported in this paper.